\documentclass[aps,prl,twocolumn,superscriptaddress,longbibliography]{revtex4-1}

\usepackage{graphicx}
\usepackage{dcolumn}
\usepackage{bm}
\usepackage{color}
\usepackage[colorlinks=true,allcolors=blue,breaklinks=true]{hyperref}
\usepackage{amsmath} 
\usepackage{rotating}
\usepackage{gensymb}
\usepackage{siunitx}
\usepackage{epstopdf}
\usepackage{soul}
\usepackage[version=4]{mhchem}

\begin{document}

\title{Direct evidence of stress-induced chain proximity in a macromolecular complex}

\author{Steven van Kesteren}
\affiliation{Physical Chemistry and Soft Matter, Wageningen University \& Research, Stippeneng 4, 6708 WE Wageningen, The Netherlands}
\affiliation{Biophysics, Wageningen University \& Research, Stippeneng 4, 6708 WE Wageningen, The Netherlands}

\author{Tatiana Nikolaeva}
\affiliation{Biophysics, Wageningen University \& Research, Stippeneng 4, 6708 WE Wageningen, The Netherlands}

\author{Henk Van As}
\affiliation{Biophysics, Wageningen University \& Research, Stippeneng 4, 6708 WE Wageningen, The Netherlands}
\email[*]{henk.vanas@wur.nl}

\author{Joshua A. Dijksman}
\affiliation{Physical Chemistry and Soft Matter, Wageningen University \& Research, Stippeneng 4, 6708 WE Wageningen, The Netherlands}
\email[*]{joshua.dijksman@wur.nl}


\begin{abstract}
The mechanical properties of many supramolecular materials are often determined by non-covalent interactions that arise from an interplay between chemical composition and molecular microstructural organization. The reversible nature of non-covalent interactions gives supramolecular materials responsive properties that are otherwise difficult to obtain, such as becoming rigid as a response to mechanical stress. How exactly non-covalent interactions emerge from microstructure, how they might change in response to applied force or deformation is not understood. Here we combine Nuclear Magnetic Resonance (NMR) and rheology  to directly probe the role of chain proximity in polymer complexes. We observe an increase in chain proximity in response to imposed flow, which we hypothesize to originate from enhanced hydrogen bonding. The chain proximity is directly correlated to rod climbing and shear banding. Flow persists only when applied stresses are low, suggesting a stress-induced thickening mechanism. We verify that hydrogen bond disruptors can turn off both the non-trivial flow behavior and the spectroscopic evidence of chain proximity. The combined rheo-NMR approach shows that it is possible to directly observe the molecular origins behind supramolecular mechanics, paving the way for further study into mechano-chemical properties of supramolecular materials.
\end{abstract}

\maketitle

Supramolecular materials such as hydrogels, vitrimers and other entangled or associated large molecular weight polymer mixtures can be uniquely adaptive and responsive, giving them great potential for applications~\cite{RN15, RN42,RN55}. The adaptive and responsive properties of these materials originate from the conformational flexibility of the molecular structure, which allows formation and dissociation of non-covalent bonds between and among molecular subunits. In general, combining such dynamic bonds in a molecular structure that is also geometrically flexible gives a huge range of possibilities to engineer supramolecular systems that can adapt and respond as desired to external stimuli~\cite{RN67, RN3, RN62, RN63}.
An interesting class of supramolecular materials are supramolecular polymer networks, in which polymers chains are cross-linked via non-covalent bonds to form reversible networks. Depending on the strength of the non-covalent cross-links and the amount of cross-linking, these materials can range from soft glue-like pastes to rubbery hydrogels~\cite{RN30}. Interestingly, these supramolecular networks have unique responses to mechanical stress: they can be stretchy yet tough, become rigid if exposed to stress~\cite{RN3,RN56}, or have reversible memory of their previous shape~\cite{RN15, RN3, RN4}. It is thought that these responses depend on the type, organization and number of non-covalent bonds, but the molecular mechanisms are not clear~\cite{RN62,RN5}. 
To understand and apply the interaction of conformational flexibility and non-covalent bond dynamics, it is necessary to have a direct view on both binding mechanisms and conformation during the act of deforming or stressing the system. Molecular conformations and (non-)covalent bonding can be readily observed by the broad toolkit of techniques that Nuclear Magnetic Resonance (NMR) offers, while mechanical testing of materials is the domain of rheology. It is therefore natural to combine these two methods in the search for optimal material design and characteristics in supramolecular mechanics.
Indeed, here we use the rheo-NMR combination to directly show that chain proximity can be ``turned on'' by flow in a polymer mixture. We show that the onset of chain proximity as measured with NMR is correlated with a slew of non-trivial flow properties of the polymer mixture, including rod-climbing and shear-thickening. To show this, we use proximity-sensitive NMR-spectroscopy techniques to study the non-covalent interactions at work in a sheared polymer mixture composed of two types of polymer chains between which hydrogen bonds act as sticky bonds. The importance of hydrogen bonding in the material used in the this study is suggested by the effect of a chaotropic salt: such an additive directly affects both the chain proximity signal and the observed rheological phenomenology. The model system we use would be considered somewhat different from more strongly bonded supramolecular networks~\cite{RN31, RN32} and should not be considered as an ideal model system to probe macromolecular complex but nevertheless, our model system will show that it is both feasible and useful to directly probe the behavior of molecular geometry in response to shear stresses. Using three types of rheology measurements combined with NMR spectroscopy, velocimetry and imaging under shear gives an in-depth understanding of the viscoelastic behavior of polymer mixtures and the molecular mechanisms behind their non-linear flow behavior.
To simplify the experimental system and NMR spectra, we use a mixture of polyethylene glycol (PEG) and polyacrylic acid (PAA). This mixture is known~\cite{RN32} to form a interpolymer complex at low pH, by hydrogen bonding between the protonated carboxylic acid of PAA and the ether of PEG (illustrated in Figure~\ref{fig:Fig1}A). The mixture of these polymers undergoes binodal decomposition, separating into a polymer-poor and a polymer-rich phase. The polymer-rich phase contains the interpolymer complexes composed of hydrogen bonded polymer chains~\cite{RN32}. The complex used is a clear continuous phase and is considered to be a (weakly) bonded network. We choose this specific network because the only non-covalent interaction between the different polymers will be hydrogen bonds, whose presence can be made plausible with NMR techniques. To get a reasonable viscosity and non-linear flow behavior in the complex, we can play with the lengths of the polymers; we used a mixture of two lengths of PEG to make the complex; see Figure~\ref{fig:FigS2} for a discussion.

\section{Measuring hydrogen bonds in static samples } 
\begin{figure}
\includegraphics[width=\linewidth]{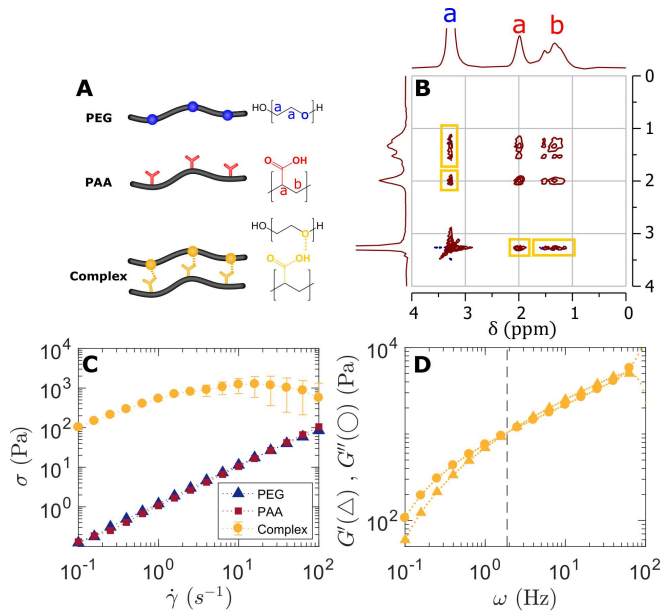}
\caption{\label{fig:Fig1} $^1$H NMR and rheological studies of the PEG-PAA complexes. (A) The schematic and molecular structures of PAA (red), PEG (blue) and the hydrogen bonded PEG-PAA complex (yellow) and the chemically unique protons labeled with lowercase letters. (B) NOESY spectrum of the PAA-PEG complex (T mixing = 300ms) with the proton-peaks identified in red and blue and the cross-correlations in yellow. (C) Flow curve of PEG 35K:300K 1:1 solution 10 \%wt. (triangles), PAA 100K solution 35 \%wt. (squares) and the PAA-PEG complex (circles) measured in a CC10-Ti concentric cylinders geometry. (D). Frequency sweep at 1 \% strain of the PAA-PEG complex. The crossover-frequency between storage (circles) and loss (triangles) moduli is indicate with a dashed line.}
\end{figure}

We use Nuclear Overhauser Effect Spectroscopy (NOESY) as 2D-NMR technique to study chain proximity between sheared polymer chains. NOESY is based on magnetization transfer between two protons via cross-relaxation, which is only possible if the nuclei are in close proximity of one another. This cross-correlation results in a cross-peak between nuclei, which are close by in space~\cite{RN38}. NOESY has already successfully been used to measure interactions in different polymers in blends and in macromolecular structures~\cite{RN28, RN29, RN14, RN37, RN26} which suggests to give a good calibration of how much hydrogen bonding is at work when the material is at rest. Our NOESY results on the PAA-PEG complexes in static conditions is shown in Figure~\ref{fig:Fig1}B, indicating strong interactions between the two polymers, most likely originating from hydrogen bond formation between the polymer chains, although this cannot be claimed with certainty: the local concentration of hydrogen bonds also plays a major role and the NOESY signal relies on a slow-motion assumption of the monomers in the chain. The NOE cross-correlations between the protons of backbones can occur via two different routes, either a direct ’A-B’ transfer from one proton to another or an indirect ’A-B-C’ transfer mediated by a third proton~\cite{RN38}. Due to the slow dynamics of macromolecules zero-quantum transitions are most probable which will result in negative cross-peaks for ’A-B’ transfers and a positive cross-peak for ’A-B-C’ transfers. The observed cross-peak is positive and therefore it is suggested the ’A-B-C’ transfer, mediated by the proton of the carboxylic acid is the main source of NOE cross-correlations~\cite{RN38,RN40, RN39}. In the end both coupling routes require the presence of a hydrogen bond and the length of a hydrogen bond is rather fixed, thus the contribution to the cross-peak intensity per bond will be similar. For this reason we will use the cross-correlation between the PAA and PEG, i.e. the NOE signal intensity, as a direct measure of the amount of hydrogen bonding between the chains. Using a binary mixture of polymers allows for the differentiation between intra- and intermolecular interactions as only intermolecular interaction between different polymers result in a NOESY cross-peak. Finally, even if the NOE signal does not directly measure hydrogen bonding, it does indicate chain proximity, also revealing important clues about the molecular geometry at play in a sheared polymer complex.

\section{Dynamic flow behavior } The PEG-PAA polymer complex has clearly obtained non-trivial flow behavior that likely originates from some form of entanglement and binding of the two different types of polymer chains. Continuous shear rheology shows that the viscosity of the complex is almost three orders of magnitude higher and depends non-linearly on the shear rate , whereas the used PEG and PAA solutions both have a Newtonian behavior (Figure~\ref{fig:Fig1}C). For the complex, the shear stress at low shear rates scales weakly sublinear with $\dot{\gamma}$; at higher shear rates it seems to level off at a plateau value. Stress saturation is even stronger in a 1~mm gap concentric cylinder system; see Fig.~\ref{fig:FigS4}. These types of flow curves cannot be fully explained by the higher concentration of polymer in the polymer complex phase only: the polymer fraction in the mix is only 17 \%wt. compared to the 10 \%wt. of the starting solutions, as measured by an evaporation study. The interactions between the polymer chains are thus likely determining the viscoelastic properties of the polymer complex. 
\begin{figure}
\includegraphics[width=\linewidth]{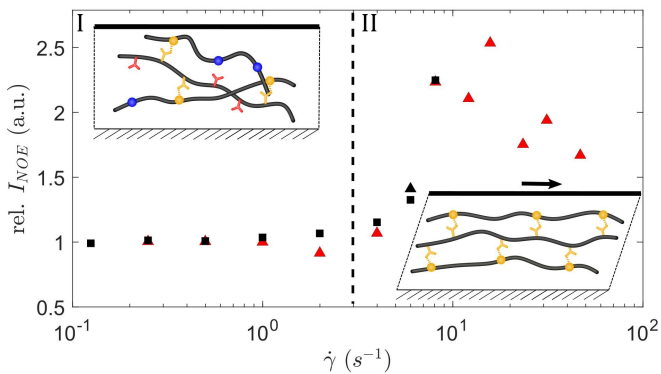}
\caption{\label{fig:Fig2} 1D-NOESY investigation measured in a 2.5mm gap in a CC-geometry. The NOE-intensity PAA-crosspeak over the PEG-peak as a measure for the number of crosslinks, normalized at $\dot{\gamma} = 0.1 s^{-1}$  (squares and triangles represent two separate measurements showing reproducibility). The dashed line shows the transition between a weak-hydrogen bonding (I) and a strong hydrogen bonding regime (II). The schematic structures illustrate a probable mechanism for the shear-induced chain alignment with likely enhanced intermolecular hydrogen bonding.}
\end{figure}

Oscillatory rheology results further strengthen the view that chain proximity is relevant for the rheology of the polymer complex. In oscillatory rheology of polymer solutions and melts the viscous (G``) and elastic components(G`) behave differently as a function of frequency compared to most other systems~\cite{RN7, RN20}. At low frequencies the polymer complex is more viscous than elastic; at higher frequencies the elastic component dominates. This behavior is most easily and customarily captured in a Maxwell model; the Maxwell model predicts a scaling of $G(\omega) \propto \omega^2$ and $G''(\omega) \propto \omega^1$~\cite{RN53}. We observe that these scalings do not hold for the PEG-PAA complex; they are closer to $G'$, $G''(\omega) \propto \omega^{0.55}$ (Figure~\ref{fig:Fig1}D). This is usually a signal of having a large range of relaxation modes in the sample, which can be caused by the large polydispersity and ``stickiness'' in the system~\cite{RN52, RN16}. We can nevertheless extract a clear timescale from the oscillatory rheology results: there is a clear crossover frequency ($\omega_{co}$) above which elastic behavior dominates viscous behavior. This transition frequency is inversely related to the slowest relaxation time of the system ($\tau = \omega_{co}^{-1}$). Frequencies below this crossover still allow the system to relax and therefore the behavior is viscous, while at frequencies above the crossover frequency the system is not able to relax during an oscillation cycle and solid-like behavior prevails. The relaxation time is an important measure for the timescales of interactions between the polymer chains, via sticky bonds or reptation~\cite{RN16, RN41, RN46}. As we will see, this crossover timescale is likely governed by the interpolymer hydrogen bonding as it can be measurably changed by adding hydrogen bond disruptors.

\section{Hydrogen bonding by shear } Having calibrated the complex’ NOESY response on a static sample and seeing non-Newtonian rheology at $\dot{\gamma} > 0 s^{-1}$, we are now in a position to observe how the NOESY signal correlates with flow. We measure 1D-NOESY spectra during continuous shear with a rheo-NMR setup, which is a setup where continuous shear can be applied to a sample within a NMR-spectrometer~\cite{RN13, RN27, RN49}. Changes in the NOE-signal can be correlated to changes in the number of bonds when shear stress is applied. A 1D-NOESY experiment measures all the NOE couplings of one specific frequency selected by a frequency-selective pulse which makes it significantly faster than a 2D-NOESY~\cite{RN35}. The frequency of the pulse in this case is the frequency of the PEG resonance signal at 2.9ppm (Fig.~\ref{fig:FigS3}), thereby observing the NOE couplings of PEG with itself and other protons, most interestingly the protons of PAA.
The relative NOE-signal as the ratio between intramolecular PEG-PEG and intermolecular PEG-PAA peaks is used to determine the amount of interpolymer hydrogen bonding (Figure~\ref{fig:Fig2}). At lower shear rates the amount of hydrogen bonds is similar to the sample at rest. The NOE signal sharply increases above a critical shear rate of $\dot{\gamma} > 3 s^{-1}$. Duplicate experiments in  the same and a narrow gap geometry (Fig.~\ref{fig:FigS4}B) show that the transition is consistently observed and the experiments are reproducible. This transition directly shows that certainly chain proximity is induced by the flow conditions; this observations constitutes the main point of the paper. Note that measurements at the highest shear rates ($\dot{\gamma} > 10 s^{-1}$) show a slight decrease in relative NOE signal. This decrease seems to coincide with a change in the rheological behavior as observed in Figure~\ref{fig:Fig1}C and~\ref{fig:Fig5}A, but the signal needs to be interpreted with care. We attribute the apparent NOE signal decrease decrease to signal acquisition limitations. An important factor to consider in these experiments is the timescale of a NOESY experiment. The cross-correlation for the NOE signal is built up over 300ms for these experiments; therefore an average amount of chain proximity over this timescale is detected and the dynamics at smaller timescales cannot be measured. This means that the average amount of chain proximity in steady state can be well detected, but as the shear rate increases, more and more hydrogen bonds live shorter than the 300ms needed to detect them, and signal intensity will decrease.

\section{Supramolecular mechanics } Why will chain proximity increase under flow? Is the transition point related to a critical rate or a critical stress? To see how this can work, it is important to take into account the conformational changes that can take place in the mixtures of flexible polymer chains in the PEG-PAA mixture. We conjecture that shear-induced bonding is due to the stretching of the polymer chains. In this picture, the NOESY data shows that a critical shear rate or stress is needed to stretch/align the polymers enough to allow for bonding sites to get close enough, before relaxation of the chain relaxes the interaction between two adjacent polymer chains~\cite{RN3, RN5}. We can further corroborate this picture by the complex flow patterns and concomitant rheological changes we observe around $\dot{\gamma} = 3 s^{-1}$.

\begin{figure}
\includegraphics[width=\linewidth]{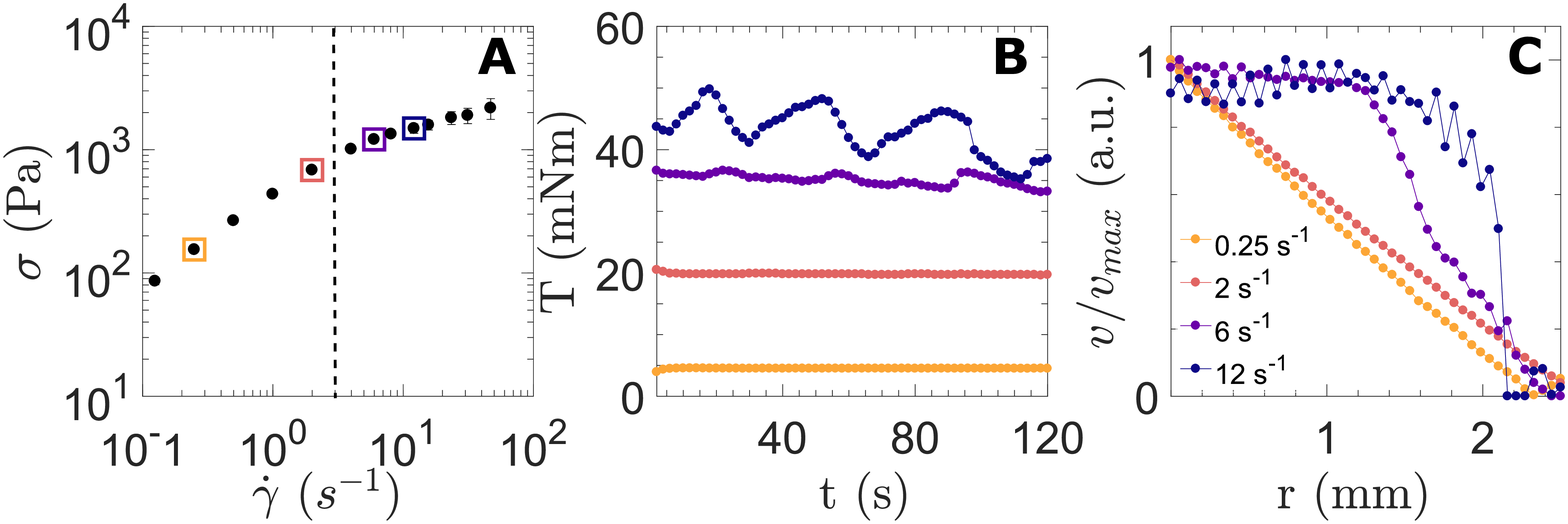}
\caption{\label{fig:Fig3}. (A) Flow curve of the PEG-PAA complex in a 2.5mm gap in a CC-geometry. (B) Torque over time at various  $\dot{\gamma}$. (C) Velocity profiles in the gap measured at various $\dot{\gamma}$ with rheo-NMR. The dashed line is the transition point between regimes.}
\end{figure}

\section{Spatial flow characterization } The flow curves measured with continuous shear rheology (Figures~\ref{fig:Fig1}C and \ref{fig:Fig3}A) show non-Newtonian behavior which cannot be directly explained by shear-induced chain proximity. With increased chain alignment under shear, one would naively expect to see an increase in viscosity due to more sticky sites such as H-bonding sites getting activated. However, the flow curve shows a decrease in shear stress compared to linear Newtonian behavior at the higher shear rates, suggesting shear thinning behavior. We resolve this apparent contradiction by examining the driving stresses and spatial flow behavior in the concentric cylinder gap during steady shear. Time-dependent torque measurements (Figure~\ref{fig:Fig3}B) show periodic fluctuations in torque occurring for shear rates only above the critical shear rate observed above. As the frequencies of these oscillations are independent of the rotation rate and its harmonics, we attribute them to a stick-slip behavior of the wall~\cite{RN34}. This stick-slip behavior suggests that at these average shear rates, the imposed stress turns the bulk material into a visco-elastic solid that yields periodically — the material is not thinning, but retains its molecular alignment and strongly elastic component and simply does not flow homegeneously anymore. The spatial velocity profiles measured with rheo-NMR ~\cite{RN13,RN27} (Figure~\ref{fig:Fig3}C) indeed clearly show the formation of shear bands at shear rates just above the transition point of increased bonding. Surprisingly, these shear bands are located at the outer static wall, indicating that most material moves almost in unison with the rotating cylinder except for a small boundary layer at the outer wall. Note that a small amount of shear remains present in the material attached to the rotating cylinder, as the velocity of the material is not proportional to $r$. The underlying mechanism here can be in line with a mechanism suggested for shear banding in polymer networks~\cite{RN36}. In this picture, polymer chains in the bulk of the material can more easily form a network than at a wall/interface, such as the outer wall. In our experiment, this picture would suggest that polymer chains are less entangled and form a less viscous material close to the outer wall, resulting in the boundary layer observed there. Note also that inside the co-rotating material, the shear rate is far below $3 s^{-1}$, suggesting further that inside this area, it is the local stress that is inducing the formation of an elastic material. Stress induced thickening is also observed in shake-gels and other shear-induced gelation mechanisms~\cite{RN3,RN62, RN63, RN52, RN66, RN69, RN65, RN64, RN61}, in which dynamic length changes of the dissolved component were previously considered responsible for the gelation~\cite{RN64, RN70}; these length changes can be a relevant microscopic mechanism in our polymer complex as well.

\begin{figure}[h]
\includegraphics[width=\linewidth]{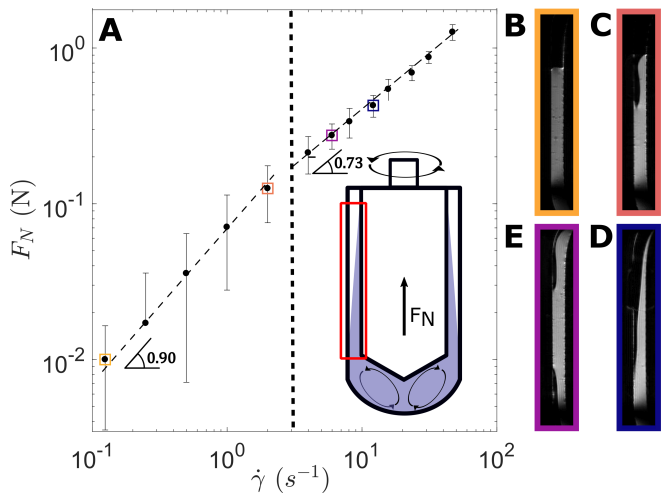}
\caption{\label{fig:Fig4}(A) Normal force on the rheometer shaft during shear of the PEG-PAA complex in a 2.5mm gap in a CC-geometry. Two different suggested shear rate scalings are indicated with the dashed lines. The illustration shows a schematic view of the CC-geometry with the proposes secondary flows and the red box is the zoomed region of the images (B-E) Zoomed MRI images of the 2.5mm gap after shearing for 10 minutes with B-E being shear rate of $0.25 s^{-1}$, $2 s^{-1}$, $6 s^{-1}$ and $12 s^{-1}$, respectively. The full images are can be found in Fig.~\ref{fig:FigS5}}
\end{figure}

\section{Normal stress measurements and MRI } Further evidence for the emergence of polymer-alignment induced complex flow patterns comes from rod-climbing observations, which was very evident during the continuous shear experiments, especially in the open cup in which the stress measurements were performed. Rod climbing also fits our picture of alignment as it can be intuitively linked to the emergence of a hoop stress, generating a pressure gradient along the radial direction, driving fluid flow up (or down) the rotating rod.
To quantify the rod climbing, we look at normal forces acting on the rotating cylinder (Figure~\ref{fig:Fig4}A) as well as MRI-images (Figure~\ref{fig:Fig4}B-E). Interestingly, the mean normal force measured on the rotating inner cylinder scales non-linearly with the shear rate; we find that the scaling appears to change to above the transition shear rate, although the difference in exponents cannot be claimed with any reasonable certainty. The same experiment has been performed in a smaller 1mm gap (see Fig.~\ref{fig:FigS6}), which resulted in a similar nonlinear shear rate dependence. We hypothesize that the conversion form the transverse rational flow into the longitudinal vortexes play a role in the shear banding observed. Specifically, these secondary flows will not only provide an upwards normal force but also inwards force perpendicular to the inner cylinder’ surface. This inwards force contributes to the separation between the strongly aligned network on the rotating cylinder and the weakly bonded layer near the wall, thereby contributing to the shear banding observed at the outside wall. MRI-images that show the distribution of material within the gap are consistent with this picture (Figure~\ref{fig:Fig4}B-E; full images in Fig.~\ref{fig:FigS5}). At lower shear rates, the material only slowly climbs up the rod, but at the higher shear rates it reaches an equilibrium distribution. Note that in all cases at least some material spans the entire gap and at these positions the velocity profiles are measured. However, the amount of material fully filling the gap thus differs between shear rates, and hence also the torque measured during shear; this partial rotor surface cover effect might partly explain the features in the flow curves, yet it is only a correction of order 1, so it cannot explain the orders of magnitude strength of the shear thinning behavior observed.\\
\section{Combining the results}
Combining the results from rheology and NMR provides a plausible hypothetical picture of how shear affects polymer complexes held together with reversible bonds. When increasing the shear rate and thus higher shear stress, the polymer chains unfold and can align, increasing the number of hydrogen bonds (as illustrated in Figure~\ref{fig:Fig2}). The shear-induced hydrogen bonding makes the polymer complexes locally more gel-like and rigid, inducing strong shear banding in the more weakly stressed outer regions of the gap. The polymer alignment also creates a normal stress on the rotor further enhancing depletion of the outer regions of the gap, and stimulating partial slipping and rod climbing. Torque oscillations can also be seen in this light: rod climbing depletes the gap and lowers the driving torque, but part of the climbing material periodically may flow back into the gap, increasing the torque again. This mechanism would however predict that the time scale of the periodicity of the torque oscillations scales the same way with shear rate as the normal force does, yet this is not observed. This slip is observed as a decrease in torque appearing as discontinuous shear thinning in steady state rheology measurements. We thus show directly that interpolymer alignment plays a critical role in the observed non-Newtonian flow behavior in this polymer system, in line with the indirect observation of the importance of hydrogen bonding in other macromolecular systems~\cite{RN67, RN62, RN56}. To further prove that the NOE-signal originates from hydrogen bonds we study the effect of breaking hydrogen bonds on the NOE-signal.

\section{Turning off hydrogen bonds } 
\begin{figure}
\includegraphics[width=\linewidth]{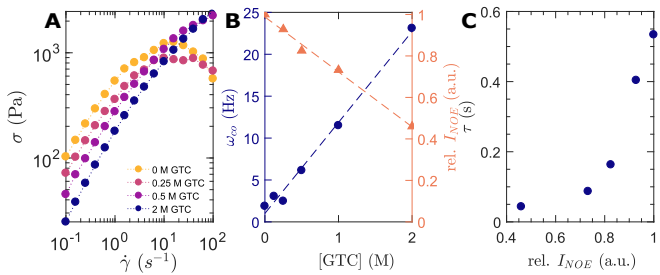}
\caption{\label{fig:Fig5} (A) Flow curve of the PEG-PAA complex at different concentrations GTC.(B)The first crossover-frequency 1 \%strain (circles) and the relative NOE-intensity (triangles) over GTC concentration. The NOE signal is normalized such that at 0 M GTC the NOE intensity is 1. (C) The characteristic relaxation times ($\tau$) over the relative NOE-intensity based on the GTC-titration. The rheological data was obtained in a CC10 setup (see experimental details)}
\end{figure}

If hydrogen bonds are responsible for the non-trivial flow behavior and NMR spectroscopy, we should be able to affect these by introducing hydrogen bond disruptors. The most common way to tune the strength of hydrogen bonds is via an increase of the temperature, however this would lead to binodal decomposition of the sample. Our method to disrupt hydrogen bonds is to use chaotropic salts because they are known from experience to break hydrogen bonds in macromolecular structures~\cite{RN63, RN56, RN54, RN33}, even though the mechanisms behind the working of chaotropic salts are still heavily debated~\cite{RN33}. We repeat NOESY and rheology tests on PEG-PAA mixtures with the well know chaotropic salt guandinium thiocyanide (GTC). We expect that GTC brings changes in the viscoelastic properties and NOE-signal. Indeed, the viscoelastic properties of the PEG-PAA mixture change when adding GTC. Steady state flow curves at different GTC concentration show that an increasing concentration of chaotropic salt makes the behavior of the polymer complex more resemble a Newtonian fluid (Figure~\ref{fig:Fig5}A). This trend is in agreement with previous experiments which showed the interpolymer hydrogen bonds are responsible for the non-linear rheology of macromolecular complexes~\cite{RN56, RN59, RN58}. Interestingly, the viscosity of the PAA-PEG mix at lower shear rates is also highly dependent on the GTC concentration, showing that both the system at rest and the response to stress are affected by the amount of hydrogen bonding. The crossover frequencies, $\omega_{co}$, measured with oscillatory rheology (full data shown in Fig.~\ref{fig:FigS7}), are used to calculate the slowest relaxations times of the complex ($\tau$). A clear inverse, linear relation between GTC concentration and the NOE-signal is observed, further cementing the correlation between the NOE-signal and hydrogen bonding (Figure~\ref{fig:Fig5}B; full spectra in (full data shown in Fig~\ref{fig:FigS8})).  From our data it is evident that at the highest GTC concentrations (lowest NOE signal), $\tau$ is governed by entanglement rather than hydrogen-based cross-linking: see the plateau at low NOE signal in Figure~\ref{fig:Fig5}C. Beyond a certain cross-link density, $\tau$ is governed by cross-links and $\tau$ seems to increase linearly with amount of hydrogen bonding. These observations are consistent with literature: very similar relations between cross-linking and relaxation have been observed previously in supramolecular networks~\cite{RN16}. The relation between $\tau$ and the number of reversible cross-links (Figure~\ref{fig:Fig5}B) is described by multiple models, which predict either a linear relationship or a power law~\cite{RN46, RN45, RN47}.

\section{Conclusion } We directly correlate the non-Newtonian rheological properties of a PEG-PAA polymer complex with changes in the amount of hydrogen bonding present in the polymer material. When exposing the polymer mixture to shear above a critical shear rate, we witness an increase in chain proximity but also flow intermittency and rod climbing. We conjecture that the shear induced thickening is induced by polymer alignment that makes the hydrogen bonding possible. Rheo-NMR velocity profiles evidence a local strong reduction of shear in the region close to the rotating cyclinder, and strong shear banding at the outer wall, all above the critical shear rate, suggesting that not the deformation rate but a certain critical stress is responsible for the observed gelation. This might be explained by a non-linear extensional rigidity in the polymer mechanics. The addition of a hydrogen-disrupting chaotropic salt affects both proximity effects and nontrivial flow behavior of the polymer complex, strongly suggesting it is the hydrogen bonding that is eventually responsible for the nontrivial mechanics observed. Supramolecular mechanics is thus clearly affected by the dynamic interplay of conformational changes in the entangled polymer mixture and non-covalent bonds between the polymers itself.
More broadly, the combination of multiple rheology and NMR experiments allowed us to build a complete picture on how non-covalent interactions govern the non-linear response of associating polymer complexes. Specifically exploring the influence of hydrogen bonds in non-covalently cross-linked polymers provides us with a molecular picture of reversible polymer networks under shear and helps in understanding rheology at the smallest scales. 
Note that our spectroscopic methods can be generalized to other non-covalent bonds, showing the strength of rheo-NMR methodology in uncovering the mechanical role of molecular interactions within a single experiment. This has significant advantages over doing exhaustive experiments that aim at excluding all other possible mechanisms~\cite{RN67, RN62, RN56, RN59, RN58, RN57, RN60} and is clearly more generally applicable. Of course, the use of NMR spectroscopy to obtain chemically-resolved rheology has already been advocated by Callaghan et. al.~\cite{RN13} some years ago. However, this methodology is still an under-explored area with the most notable examples being the study of \textsuperscript{13}C spectra hyaluronan, a polysaccharide, under shear which shows the breaking of hydrogen bonds and is linked to changes in the tertiary structures and subsequent shear thinning~\cite{RN19}. In a more recent example, Iwakawa et. al. have used high resolution rheo-NMR spectroscopy to study the structure of proteins under shear~\cite{RN50, RN51}. Bringing the analytical strengths of high-resolution NMR spectroscopy to the realm of rheology and material testing provides crucial insights in the structure of materials under non-equilibrium conditions.\\

The authors thank Alexey Lyulin, Carlo van Mierlo and Jasper van der Gucht for helpful discussions.

\section{Appendix A: Methods \& Materials}

\begin{figure*}[!t]
\label{FigS1}
\includegraphics[width=14cm]{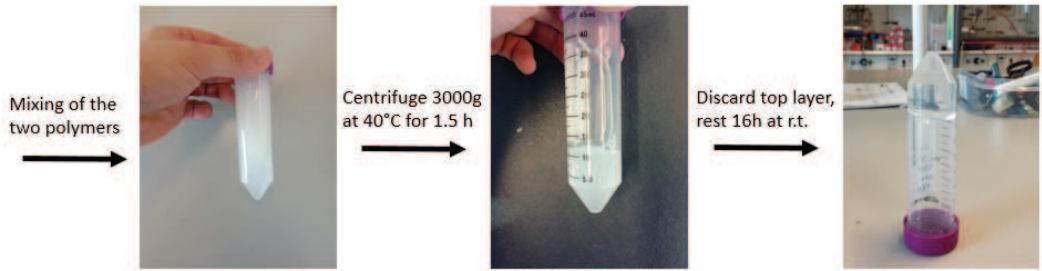}
\caption{\label{fig:FigS1} Pictures showing the steps making of the polymer complex}
\end{figure*}

\subsection{Rheology} \label{app:rheo}
The rheological experiments were performed on Anton Paar MCR301 or MCR501. Two geometries were used; initial flow curves and the GTC titration flow curve and oscillatory measurements were conducted in a commercial CC10-Ti concentric cylinders geometry with inner diameter = 10.005mm, outer diameter = 10.835mm (400 micron gap) and a length = 15.025mm. The measurements complementary to the rheo-NMR were performed in a custom-made concentric cylinder geometry with inner diameter = 17mm, outer diameter = 22mm (2.5 mm gap) and a length = 58mm. Additional checks for Fig.~\ref{fig:FigS4},\ref{fig:FigS6} were performed on a second custom CC geometry, with a 1~mm gap, inner diameter = 20mm and outer diameter = 22mm. All setups used a stationary outer cylinder and a rotating inner cylinder; the rheology geometry was also used inside the NMR for the rheo-NMR experiments~\cite{RN49}.

\subsection{NMR and rheo-NMR}
The rheo-NMR experiments were performed on a Bruker 300MHz Wide-Bore spectrometer with a MicWB40 micro-imaging probe, Avance I console and the Rheo-NMR accessories, with custom-build CC~\cite{RN49}. The stationary experiments including the GTC-series were performed on a Bruker 500MHz spectrometer with TXI probe and Avance III console. Here a 2D-NOESY sequence was used with the following parameters: TR = 2.5s, mixing time = 300ms, NA = 8, dimensions = 4096*512. The NOE data in the rheo-NMR setup (Figure 3) was obtained in the following manner: For the rheo-NMR experiments a 1D-NOESY was measured using a stimulated echo pulse sequence (Figure S3A) with TR = 1.5s, TE = 304ms, mixing time = 300ms, NA = 64, phase-cycling (see table in Fig.~\ref{fig:FigS3}) and frequency selective excitation pulse: gauss, frequency = 2.9ppm, bandwidth = 200Hz (0.67ppm). For the stationary experiments a 2D-NOESY was used with mixing time = 300ms. The magnitude of the baseline-corrected NOESY-spectra were used and ratio of the integrals of the peaks at 2.9 and 1.5ppm is used as the relative NOE-signal. This signal was normalized to zero shear = 1 for the rheo-NMR and zero GTC = 1 for the GTC experiments. More details can be found in the Supplementary Information section 2. The velocity profiles were measured using a spin double echo velocimetry~\cite{RN49} sequence (TR = 1.5s, $\Delta$ = 17.5ms, $\delta$ = 1ms, NA = 16) at the position were the MRI images showed sample to be present. The MRI images were taken with a RARE-sequence (Resolution = 256*256, FOV = 5cm*5cm, RARE-factor = 32, TE = 72ms ,TR = 1.5s, NA = 1, measuring time = 24s) with 6 slices interlaced at 30 degree angles except for image after 0.25s$^{-1}$, which was just one slice.

\subsection{Ingredients and preparation}

The poly(ethylene glycol) MW = 35KDa (PEG 35K), poly(ethylene glycol) MW = 300KDa (PEG 35K), poly(acrylic acid) MW = 100KDa (PAA 100K) and the guadinium thiocyanate (GTC), were obtained from Sigma-Aldrich. 60g PEG 35K was dissolved in 240g Milli-Q water by stirring for 16h to make clear solution of 20 \%wt. which was used as without further purification. 13.5g PEG 300K was dissolved 186.5g Milli-Q water by stirring for 16h, insolubles were remove by centrifugation at 10000g for 8h, leaving a clear 6.67 \%wt. solution. A mixture of PEG 35K 20 \%wt. and PEG 300K in a ratio 1:3 resulted in the 10 \%wt. solution of PEG 35K:300K 1:1. The PAA 100K was provided in a 35 \%wt. solution which was diluted with Milli-Q water to a 10 \%wt. solution with a pH of 1.\\
Figure~\ref{fig:FigS1} shows a visual guide through the process of making the polymer complex. The interpolymer complexes were prepared as follows. 25ml of 10 \%wt. solution of PEG 35K:300K 1:1 was mixed with 25ml of a 10 \%wt. solution of PAA 100k (pH $\sim$1). The mixture was an opaque suspension , with a pH of 1.8 containing some aggregates, most interpolymer complexes are reported as turbid suspensions of small aggregates of complexed polymer in a polymer-poor phase~\cite{RN32}. This solution is centrifuged for 1.5h at 3000 g and 40 degrees C which separates the polymer-poor and the polymer-rich phase~\cite{RN32}. The elevated temperature of centrifuging employs the phase behavior of the sample for separation, as at higher temperatures the phases mix less well with water than at room temperature, which allows the polymer sample to adhere as a continuous phase at room temperature. The polymer-poor phase is discarded and the final complex is formed by letting the sample rest at room temperature for 16 hours. The final PAA-PEG complex phase was a clear, viscous, gel-like substance with pH $\sim$1.5; see also Figure~\ref{fig:FigS1}. The polymer concentration of this bottom phase was determined to be 17 \% weight by rotary evaporation. For the experiments with 2D-NOESY (initial measurements and GTC titration), a  10 \% PAA solution was made by diluting with Milli-Q water containing 20 \% $^2$H$_2$O. This $^2$H$_2$O is needed for the deuterium lock. For the GTC titration 0.125, 0.25, 0.5, 1 and 2mmol GTC was added to 1g ($\sim$ 1 ml) of PAA:PEG complex to make the 0.125, 0.25, 0.5, 1 and 2M solutions, respectively. The pH of any solution did not exceed 2, well below the pI of PAA ($\sim$4.5).  

We choose this particular combination of PEG-PAA polymers because of their simple 1H-NMR spectra with good separation between the different protons, and a strong detectable NOE-signal which directly evidences chain proximity. We tried different potential hydrogen bonding homo-polymer combinations, such as polyvinylalcohol and polyethylenimine, which did not show the desired spectroscopic properties.  Furthermore, by tuning the length of our PEG-chains some practical handling difficulties, like zero-shear viscosity, equilibration and stickiness, were managed. 

\begin{figure}[!h]
\label{FigS2}
\includegraphics[width=\linewidth]{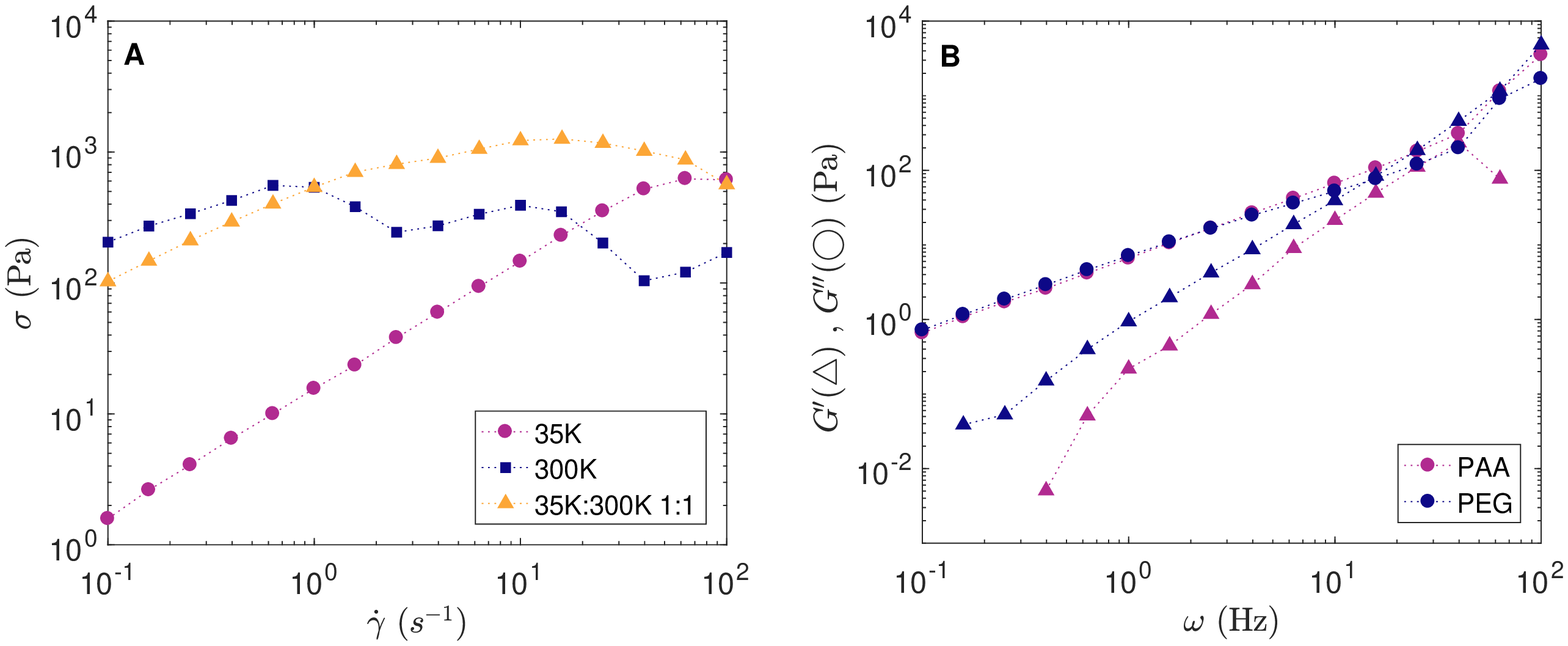}
\caption{\label{fig:FigS2} (A) Flow curves of the polymer complexes made with PAA 100K and PEG 35K (circles), PEG 300K (squares) and the 35K:300K 1:1 ratio (triangles) (B) Frequency sweep at 1 \% strain of the PAA 100K solution (purple) and PEG 35K:300K 1:1 solution (blue), with triangles indicating the storage modulus and circles the loss modulus.}
\end{figure}

The reason for using the particular mixture of PEG 35K and PEG 300K is shown in Figure \ref{fig:FigS2}A. An complex of PAA 100K and PEG 35K shows mostly Newtonian behavior with only some shear rate dependence at very high shear rates. The complex of PEG 300K shows the long polymer chains make the sample too viscous to use in the rheo-NMR setup and the non-Newtonian regime starts at too low shear rates. The mixture of PEG 300K and PEG 35K in 1 to 1 ratio showed the best "middle-ground" properties with a workable viscosity and a clear and reproducible transition to a non-Newtonian regime at intermediate shear rates.

Figure \ref{fig:FigS2}B shows the frequency sweep of the individual PAA 100K and PEG 35K:300K solutions. The PEG-solution has a crossover frequency at 17 Hz while the PAA-solution shows no crossover. Furthermore the low polydispersity PAA 100K shows rheological behavior that fits a single mode Maxwell model quite well, i.e. G' = $\omega^2$ and G'' = $\omega^1$. The high polydispersity mixture of PEG 35K:300K deviates significantly from these scalings, as expected for a mixture of polymers in which multiple relaxation times are present. 

\section{Appendix B: NOESY measurements under shear}

\begin{figure}[!hb]
\label{FigS3}
\includegraphics[width=\linewidth]{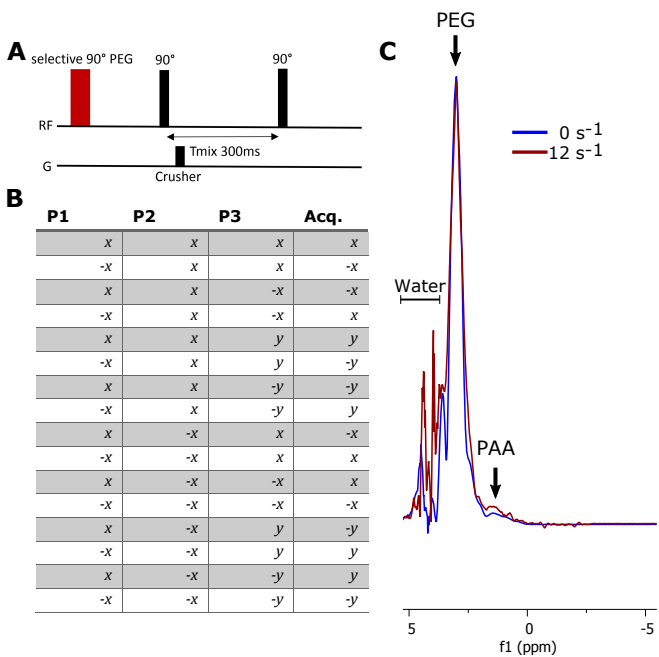}
\caption{\label{fig:FigS3} (A) Schematic overview of the pulse sequences used for the 1D-NOESY. (B) Table of the phase cycling used to select the correct coherences, where P1, P2 and P3 stand for the phase of the first, second and third pulse respectively and Acq. stands for the receiver phase. (C) Examples of the 1D-NOESY spectra at 0 and 12 $s^{-1}$ in a 2.5 mm gap normalized at the PEG peak, showing the increase in NOE-signal due to shear-induced bonding. Note that the water peak is not entirely canceled out due to imperfections of phase cycling.}
\end{figure}

The pulse sequence and phase cycle for the 1D-NOESY are show in Figure \ref{fig:FigS3}A and \ref{fig:FigS3}B, respectively. The pulse sequence was adapted version of the Stimulated Echo Acqustion Mode (STEAM) from Paravision 5.1 with a frequency selective pulse, different phase cycle and a additional mixing time delay. The red selective pulse of Figure \ref{fig:FigS3}A was a frequency selective excitation pulse: gauss, frequency = 2.9 ppm, bandwidth = 200 Hz (0.67 ppm). The 1D-NOESY was tested by varying the mixing times and seeing the NOE-signal disappear at short mixing times. The rheo-NOESY experiment (Figure 3) was done in the following manner: The system was wobbed, shimmed and the pulses were calibrated. The selective pulse of the 1D-NOESY was tested and then rheo-cell the system was spun for about a minute for equilibration. After equilibration a 1D-NOESY was obtained, followed by a velocity profile, then the rotation was stopped an immediately a MRI-image was taken. This process was repeated for every shear rate.     

The data was processed as follows: All processing of the 1D-NOESY spectra was done using Matlab 2017. The raw NOESY spectra were loaded, Fourier transformed and the magnitude is used to ensure no biases in phasing were present. The spectra were baseline corrected by subtracting the average of a part of the spectrum without features (between 7 and 7.5 ppm). Examples of these spectra are shown in Figure \ref{fig:FigS3}C. The integral of the region between 1.40 and 1.60 ppm was divided by the region between 2.80 and 3.00 ppm and these ratios were normalized to zero shear = 1. This results in the normalized NOE-signal used in the experiments. The 2D-NOESY spectra were processed with MestReNova. The spectra were loaded, automatically phased and baseline corrected with the built-in functions. Then the same PAA and PEG regions (1.40 to 1.60 ppm and 2.80 to 3.00 ppm, respectively) were integrated and these integrals were divided and normalized to zero GTC concentration. 

\begin{figure}[!h]
\label{FigS4}
\includegraphics[width=\linewidth]{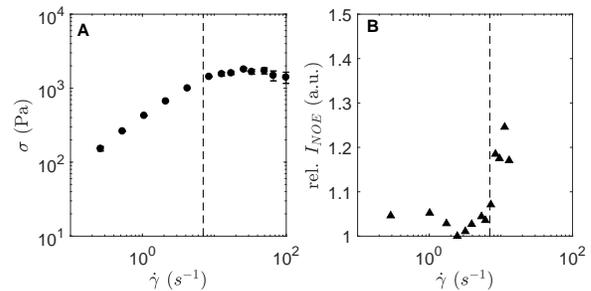}
\caption{\label{fig:FigS4}(A) Flow curve of the PEG-PAA complex in a 1.0 mm gap in a CC geometry. Note that the rate at which the shear thinning sets in is slightly higher than in Fig.~\ref{fig:Fig1}, but the stress level is similar. (B) The NOE-intensity PAA-crosspeak over the PEG-peak as a measure for the number of cross-links, normalized that $\dot{\gamma}_0$ = 1 (triangles). }
\end{figure}

\begin{figure}[!h]
\label{FigS6}
\includegraphics[width=\linewidth]{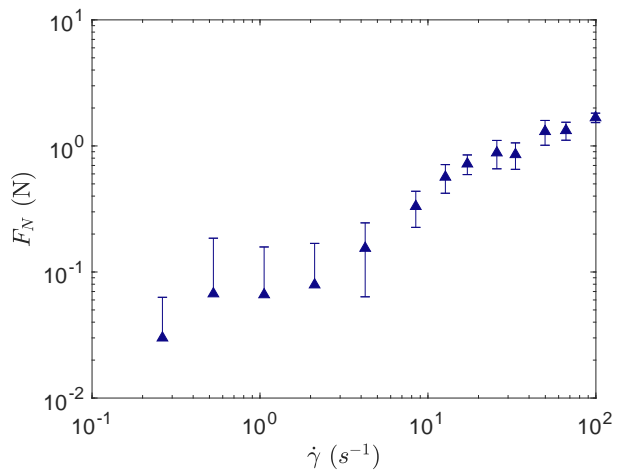}
\caption{\label{fig:FigS6} Normal force on the rheometer shaft during shear for the 1.0 mm gap.}
\end{figure}

\begin{sidewaysfigure}
[!ht]
\label{FigS5}
\includegraphics[width=\textwidth,viewport=200 250 1300 500,clip]{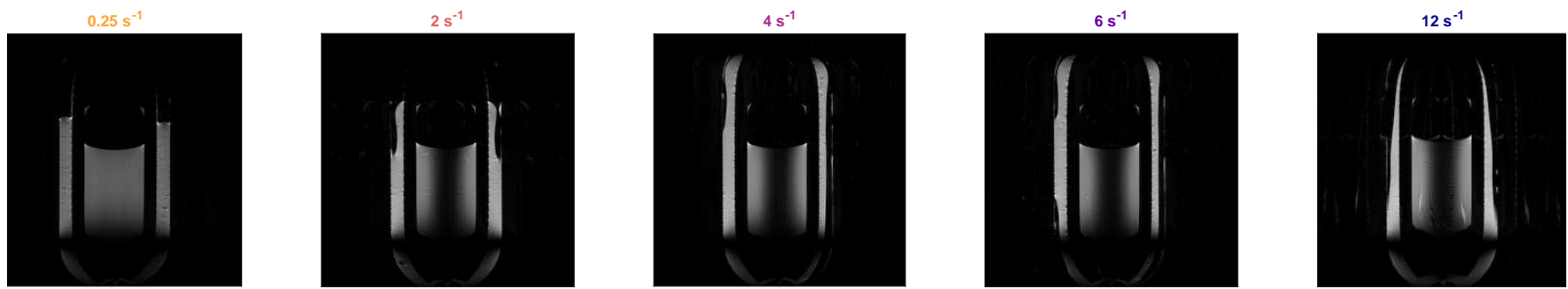}
\caption{\label{fig:FigS5} MRI images of the rheo-NMR cell with 2.5 mm gap after shearing for 10 minutes at various rates. (Sequence RARE 256*256, FOV = 5cm*5cm, RARE-factor = 32, TE = 72 ms ,TR = 1.5 s, NA = 1, measuring time = 24 s)}
\end{sidewaysfigure}

\begin{figure*}[!h]
\label{FigS7}
\includegraphics[width=\linewidth]{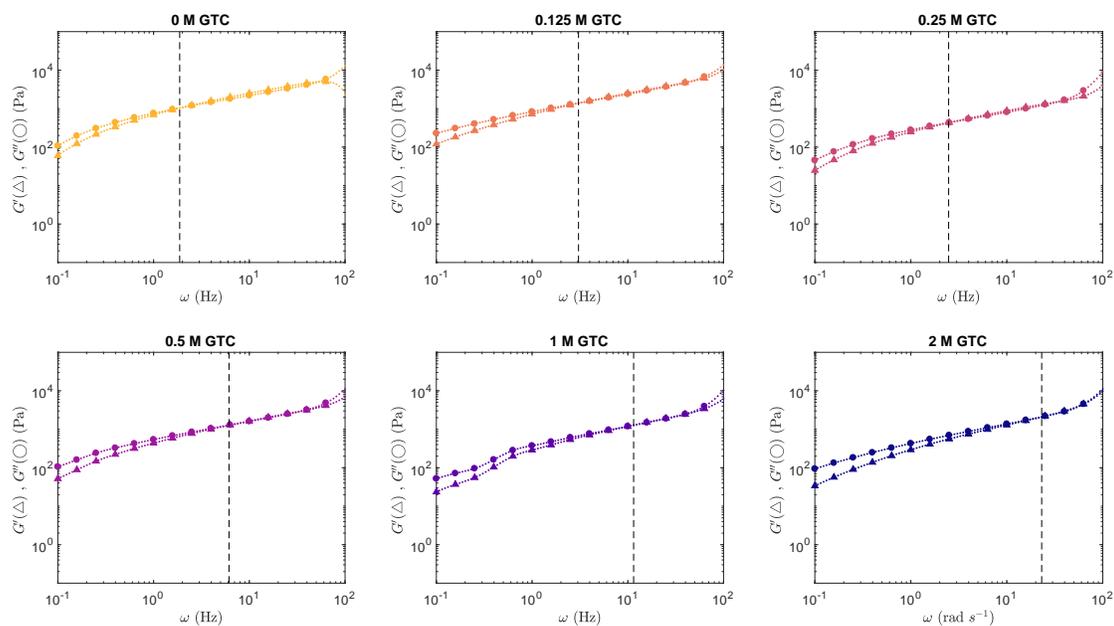}
\caption{\label{fig:FigS7}Frequency sweep at 1 \% strain of the PAA-PEG complex with increasing concentrations of GTC. The crossover-frequencies between storage (circles) and loss (triangles) moduli are indicated with a dotted line. These crossover-frequencies($\omega_co$) are used in Figure 5B.}
\end{figure*}

\begin{figure*}[!h]
\label{FigS8}
\includegraphics[width=\linewidth]{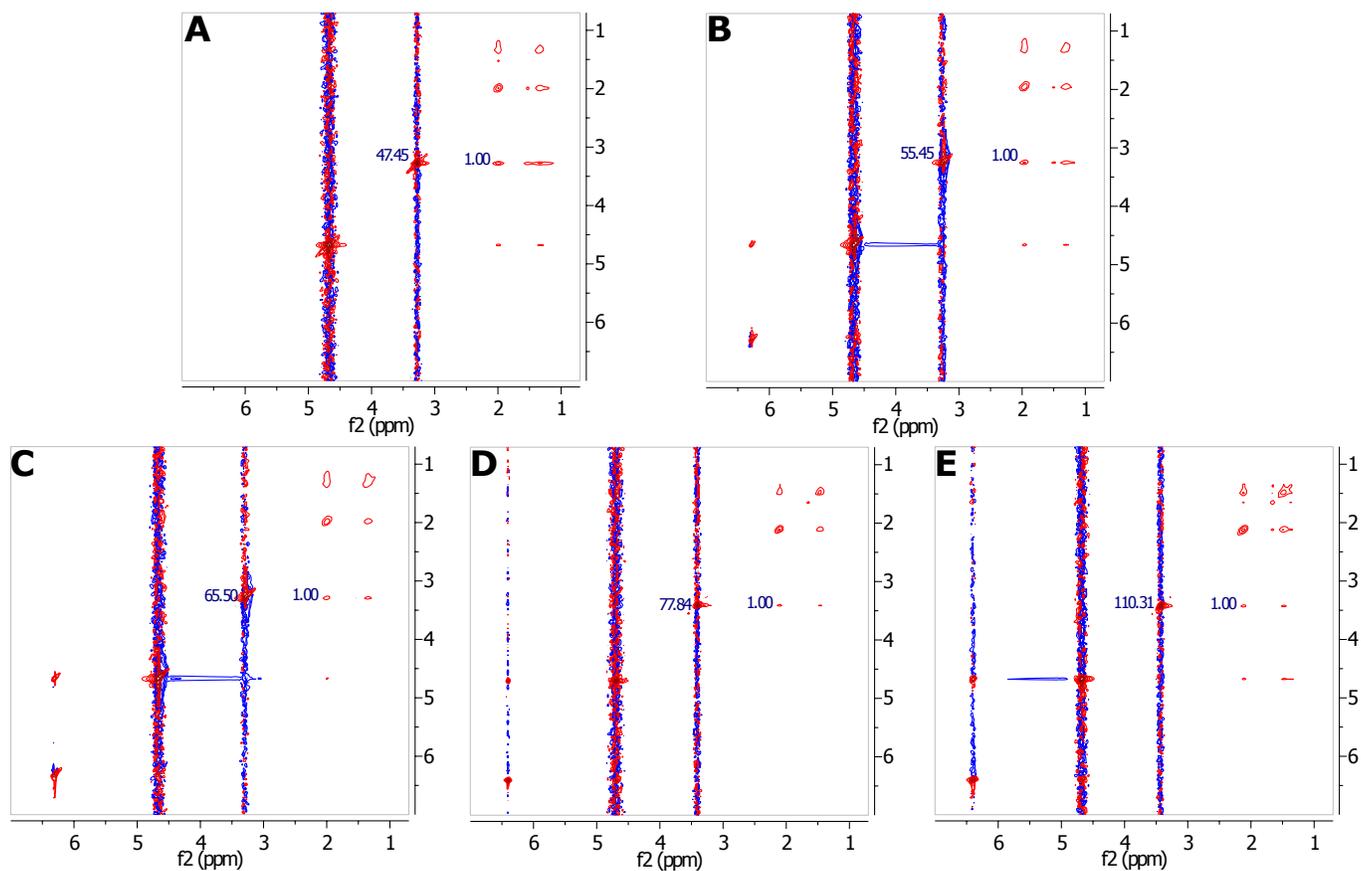}
\caption{\label{fig:FigS8} NOESY spectra including the integrals at GTC concentrations of 0 M, 0.25 M, 0.5 M, 1.0 M and 2.0 M of A-E,respectively. These integrals to calculate the relative NOE-signal are used in Figure 5B.  }
\end{figure*}

We are aware that there are other explanations for changes in NOE-signal besides hydrogen bonding, such as changes in mobility or motion artifacts. However, these alternative explanations would likely affect the NOE-signal at all shear rates and would not coincide with the other results like the shear banding. Therefore we are confident our linking NOE-signal to hydrogen bonding is a valid approach, especially combined with the GTC experiments. Also the 1.0 mm gap (Figure \ref{fig:FigS4}B) shows increase in NOE-signal at a critical shear rate at although less pronounced as in the 2.5 mm gap. This increase in NOE-signal again coincides with a decrease in shear stress which was determined to be a sign of shear banding for this system. The smaller increase in signal in a 1.0 mm gap compared to a 2.5 mm gap expected as the non-bonded boundary layer near the wall is relatively larger for a smaller gap size and the hydrogen bonded layer is thus smaller. One additional check that could be performed to verify the role of the carboxylic proton is to use and retain pure \ce{D_2O} as a solvent for the polymers in the entire preparation phase of the sample.     

\section{Appendix C: supplementary rheology and NMR data}

\clearpage


%

\end{document}